\documentclass{aa}
\usepackage{amssymb}
\usepackage{psfig}
\usepackage{graphicx}
\usepackage{mathptm}
\begin{document}
\title{Gravitational tidal effects on galactic open clusters\thanks{Plate
scanning was done at the Centre d'Analyse des Images (CAI)  with 
{\textsf{\textsl{M.A.M.A.}}} 
({\sl Machine Automatique \`a Mesurer pour l'Astronomie}), a facility located at
the Observatoire de Paris, developed and operated by INSU (Institut National 
des Sciences de l'Univers, CNRS). Web site 
{\tt http://dsmama.obspm.fr}}} 

   \author{G. Bergond\inst{1,2},
        S. Leon \inst{3},
         \and
          J. Guibert\inst{1,2}
        }

   \offprints{G. Bergond ({\tt gbergond@eso.org})}
   \institute{Centre d'Analyse des Images, 
Observatoire de Paris, 77 Avenue Denfert-Rochereau,
        F--75014 Paris, France
		\and
DASGAL/UMR 8633, Observatoire de Paris-Meudon, 5 Pl. Janssen, 
F--92195 Meudon cedex, France
                \and
I. Physikalisches Institut, Universit\"at zu K\"oln, 
Z\"ulpicher Stra\ss e 77, D--50937 K\"oln, Germany}
   \date{Received XX; accepted XX}

   \abstract{
We have investigated the 2--D stellar distribution in the 
outer parts of three nearby open clusters:
\object{NGC~2287} ($\equiv$\,M41), \object{NGC~2516}, 
and \object{NGC~2548} ($\equiv$\,M48).
Wide-field star counts have been performed in two colours on pairs of
digitized ESO and SRC Schmidt plates,  allowing us to select 
likely cluster members in the colour-magnitude diagrams. 
Cluster tidal extensions were emphasized using
a wavelet transform. Taking into account observational biases, 
namely the galaxy clustering and differential extinction in the Galaxy,
we have associated these stellar overdensities with real open cluster
structures stretched by the galactic gravitational field. As predicted by
theory and simulations,  and despite observational limitations,
we detected a general elongated (prolate)
shape in a direction parallel to the galactic Plane, combined
with tidal tails extended perpendicularly to it. This geometry
is due both to the static galactic tidal field and the heating 
up of the stellar system when crossing the Disk. The time
varying tidal field  will deeply affect the cluster dynamical
evolution, and we emphasize the importance 
of adiabatic heating during the Disk-shocking. 
In the case of NGC~2548, our dating of the last shocking with the Plane (based
on a tidal clump) is consistent with its velocity.
During the 10--20 $Z$-oscillations experienced by a cluster before
its dissolution in the Galaxy, crossings through the galactic Disk 
contribute to at least 15\% of the total mass loss.
Using  recent age estimations published for open clusters, we find a
destruction time-scale of about 600 Myr for clusters 
in the solar neighbourhood. 
}
\maketitle
\section{Introduction}

Galactic open clusters (OCs) have long been shown to be
extremely valuable laboratories for many domains of astronomy. In
particular, observations of their spatial structures can help to
constrain dynamical models and \mbox{$N$-body} simulations of small
($N \approx$ 10$^3$) stellar systems. Extended over some parsecs, OCs 
are made up from many dozens to a few thousands stars.
Their sparseness makes them short-lived stellar concentrations.

In this manner, most of galactic OCs evaporate 
entirely in some 10$^8$ years: on the 1200 or so objects of this
type in the catalogue of open cluster data compiled by Lyng\aa\
(\cite{lynga87}), only about 70 are known to be older than 1 Gyr. 
Indeed, as a result
of 2-body relaxation, several stars can acquire  positive energy
during the strongest (nearest) interactions with other members,
and then leave the cluster which slowly vanishes. Stellar evolution
(mass loss) and encounters with giant molecular 
clouds (GMCs, see Wielen \cite{wielen91}) also contribute  
to noticeably reduce the stellar system lifetime. In addition, the galactic potential 
plays a major role in the disruption process by its regular
gravitational harassment, through tidal disruption. 

As a consequence, only the initially
richest clusters (which are more gravitationally bound)
and those situated at large galactic radii (where the probability
of encounter with a GMC is lower) can live a few Gyr.
De la Fuente Marcos (\cite{fuente98}) has shown that
our Galaxy should host several hundreds of thousands
of open cluster remnants which have been disrupted
by the Galaxy and by their own dynamical evolution
(evaporation). Bica et al. (\cite{bica01}) indeed observe several
dissolving star cluster candidates. 

Orbits of OCs being quasi-circular, with $Z$-oscillations
of small ($\lesssim$ 1 kpc) amplitude, their trajectory
includes many passages through the Disk: each of these
gravitational shocks heat up and compress the cluster which then takes a 
prolate shape flattened to maximum at $Z = 0$ (Leon \cite{leon98}). 
Repeated disk-shockings speed up the disruption, as studied in globular 
clusters (Combes et al. \cite{combes99}) and after each passage
through the Plane, members rejected in the halo of the system are
stripped out by the gravitational field of the whole
Galaxy. These ejected stars will form a tidal tail which 
extends  far from the cluster inner regions. Structural
studies of the surroundings of OCs can thus provide much 
information on the past and present dynamical evolution of the cluster.

Terlevich (\cite{terlevich87}) and de la Fuente Marcos (\cite{fuente97} and
references therein) performed realistic simulations
of open cluster evolution. From the observational point of view, 
Grillmair et al. (\cite{grillmair95}) and Leon et al. (\cite{leon00})
 detected important tidal tails extensions around galactic {\em globular}
clusters using a wide-field star count analysis. Leon et al. (\cite{leon99})
also found noticeable tidal tails around several binary star clusters 
in the Large Magellanic Cloud.

The main aim of this study is to detect such tidal tail signatures
in the overall cluster structure around selected galactic OCs, by performing
multicolour, wide-field star counts on Schmidt photographic plates.
First results are presented in Bergond et al. (\cite{bergond01}).
In Sect. 2 we discuss the selection of three open clusters.
Section 3 presents the data analysis, and a brief discussion 
on our observational limitations.
In Sect. 4 we focus on each of the three  observed OCs.
Next, in Sect. 5, we discuss these results together 
with numerical simulations, and their general 
implications for galactic open clusters,
before summarizing this work in the last section.

\section{Open cluster sample}

It is expected that tidal tails are very weak stellar overdensities, and
to better detect them it is necessary to reject at maximum
areas with strong background variations due to a differential interstellar
extinction. A hole of extinction can indeed artificially enhance the star-count
in this more transparent area, giving the impression of a higher density of 
stars which we could mistake for a tail. For this reason, we have made a first 
selection, restricting our choice to the few OCs which present 
$b^{\rm{II}} >$~10\degr\
(see Janes \& Adler \cite{janes82} for the galactic distribution of OCs).
Such high latitudes allow us to avoid the major part of dust clouds which are
generally well concentrated along the Plane $Z =$ 0,
 with a mean scale height of 160$\pm$20 pc 
(Pandey \& Mahra \cite{pandeymahra87}).

\begin{figure}
\resizebox{\hsize}{!}{\includegraphics{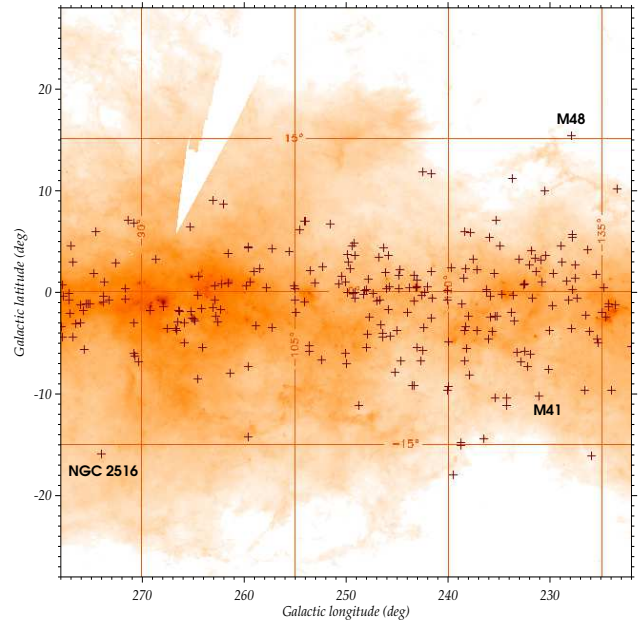}}
\caption{{\tt Skyview} map of {\sl IRAS} 
dust emission at 100 $\mu$m in the regions going from $(\ell,b) =$ 
(222\degr,$-$28\degr) to $(\ell, b) =$ (278\degr,+28\degr). 
All the known open clusters in the Lyng\aa\ catalogue 
are shown as small crosses (+).
The three selected objects are indicated.
They avoid strong and differential extinction areas.}
\label{montage}
\end{figure}

However, the extinction within the Plane is highly variable from one direction
to another, and there are windows of transparency, especially between $\ell =$
210\degr\ and $\ell =$ 240\degr\ (Chen et al. \cite{chen98}).
The whole map of
dust emission (as traced by {\sl IRAS} 100 $\mu$m) 
in these directions is visible in Fig.~\ref{montage}
where we have overplotted all the known OCs in the Lyng\aa\ (\cite{lynga87})
catalogue.

We had yet to select  relatively rich clusters in order to perform a
reliable star-count analysis. Very poor OCs do not offer a sufficient 
contrast with respect to the field stars, and may be only {\it asterisms}, that is 
apparent concentrations of stars due to a pure perspective effect 
(e.g., Baumgardt \cite{baumgardt98}).
At last, the cluster must have the appropriate dimensions: too small clusters
are difficult to treat with the typical scale on Schmidt plates (67\farcs1/mm)
due to limitations in resolution of emulsions and diffusion 
problems in crowded areas, whereas really too wide clusters
like Melotte 111 (Odenkirchen et al. \cite{odenkirchen98})
cannot be studied on one single $\approx 30$ deg$^2$ plate. 

We finally selected as the most interesting
 candidates 3 rich, bright and low-absorbed OCs:
NGC~2287 ($\equiv$\, M41), NGC~2516, and NGC~2548 ($\equiv$ M48). 
Table~\ref{plates} presents the pairs of photographic
plates we scanned for this study and 
Table~\ref{taboc} summarizes the main 
properties of these OCs.

\begin{table}
\caption[]{Scanned Schmidt plates encompassing the clusters.}
\label{plates}
\begin{tabular}{llllll}
\hline
Cluster & Plate & Emuls. & Filter & Exp. & Epoch\\
NGC & ID \# & Kodak & used & min & \\
\hline
2287 & SRC557 & IIIaJ & GG395 & 60 & 1979.898\\
2287 & ESO557 & IIIaF & RG630 & 60 & 1985.930\\
\hline
2516 & E11031 & IIIaJ & GG385 & 15 & 1994.117\\
2516 & E11190 & IIIaF & RG630 & 15 & 1994.444\\
\hline
2548 & SRC775 & IIIaJ & GG395 & 60 & 1986.246\\
2548 & SRC775 & IIIaF  & OG590 & 70 & 1985.057\\ 
\hline
2548 & SRC776 & IIIaJ & GG395& 60 & 1983.192\\
2548 & SRC776 & IIIaF & OG590& 70 & 1985.953\\
\hline
\end{tabular}
\end{table}
\begin{table*}
\caption[]{Main characteristics of the three studied open clusters 
 as extracted from the {\sl Base des Amas} (see Mermilliod \cite{mermilliod95})
{\tt http://obswww.unige.ch/webda/}, 
except for the masses (particularly uncertain),
which come from Bruch \& Sanders (\cite{bruch83}) for NGC 2287, or from
Pandey et al. (\cite{pandey87}) for NGC 2516. Some values showing important
discrepancies in the literature are specified in the note below the table.}
\label{taboc}
\begin{tabular}{ccccccccccccc}
\hline
Cluster & $\alpha$ & $\delta$ & $\ell$ & $b$ & Dist & $Z$ & 
Rad. & Mass & Age & Spect.-type &  [Fe/H] & $E_{B-V}$\\
name & $^{\rm{J2000.0}}$ & $^{\rm{J2000.0}}$ 
& $^{\rm{II}}$ & $^{\rm{II}}$ &pc & pc & 
pc & M$_{\sun}$ & Myr & on turn-off & dex$^{\ddag}$ & mean\\
\hline
NGC 2287 & 06\fh46\fm9 & $-$20\degr44$'$ & 231\fdg10 & $-$10\fdg23 & 693 
&$-$123& 4.1 & 294 & 243 & B5 & $+$0.04 & 0.03\\
NGC 2516 & 07\fh58\fm3 & $-$60\degr52$'$ & 273\fdg94 & $-$15\fdg88 & 409 
&$-$112& 1.9 & 170 & 113$^{\dag}$ & B3 & $+$0.06 & 0.10\\
NGC 2548 & 08\fh13\fm8 & $-$05\degr48$'$ & 227\fdg92 & 
$+$15\fdg37 & 769 &$+$204&
4.8 & N/A & 360 & A0 & $+$0.08 & 0.03\\
\hline
\end{tabular}
\begin{list}{}{}
\item[$^{\dag}$] Meynet et al. (\cite{meynet93}) fit NGC 2516 with
an older isochrone at 140 Myr.
\item[$^{\ddag}$] 
For NGC 2516, Jeffries et al.
(\cite{jeffries97}) proposed a value of [Fe/H] $\simeq$ $-$0.32.
\end{list}
\end{table*}


\section{Data analysis}
All the plates were digitized with a resolution of 10~$\mu$m with the 
{\it Machine Automatique \`a Mesurer pour l'Astronomie} ({\textsl{\textsf{MAMA}}}, 
see Berger et al. \cite{berger91}). Source extraction was done
using the {\sf SExtractor} software (Bertin \& Arnouts \cite{bertin96}),
at a 3$\sigma$-level for each pair of plates. Once the
astrometry was performed in each colour, local software was used for the 
cross-correlation of the $B$ and $R$ catalogues. 

This allows us to build a colour-magnitude diagram
(CMD), in {\it instrumental} magnitudes, for the whole plate.
Next, we adopt the method described by Grillmair et al. (\cite{grillmair95}),
which consists of selecting
regions in the CMD where the ``cluster {\it vs.} field contrast'' is 
important. For this purpose, we use a so-called colour-magnitude S/N
function $\overline{s}(i,j)$ defined by subdividing the CMD in multiple cells
(50 $\times$ 50 bins here,  i.e. about 0.1 mag both in $B$ and $B-R$). 
We refer to the papers of Leon et al. (\cite{leon99}, \cite{leon00}) 
for  details of the application of the method.
We then retain only cells where the number of cluster members 
with respect to the field stars is the highest, by 
defining a threshold we choose as a compromise
between significant star-count (in order to prevent background fluctuations)
 and  reduced contamination by field stars.

In the case of relatively nearby clusters, it is possible 
to use a direct ``cut'' in magnitude in the whole star count,
selecting only the brightest sources.
Most of weak field stars are then 
directly eliminated, whereas only the bottom of the OC main sequence 
(that is, most of late-type dwarf members) is missing.
Keeping only bright OC members to minimize pollution also obviously decreases  
 the number of detected 
 stripped stars from the cluster, which are preferentially 
 low mass stars, as noted both from observations (Leon et
al.~\cite{leon00}) and simulations (e.g., Combes et al.~\cite{combes99}),
but the S/N of the tidal tails relative to the background/foreground stars
will be higher due to the richness of the
three  selected clusters.

\begin{figure}
\resizebox{\hsize}{!}{\includegraphics{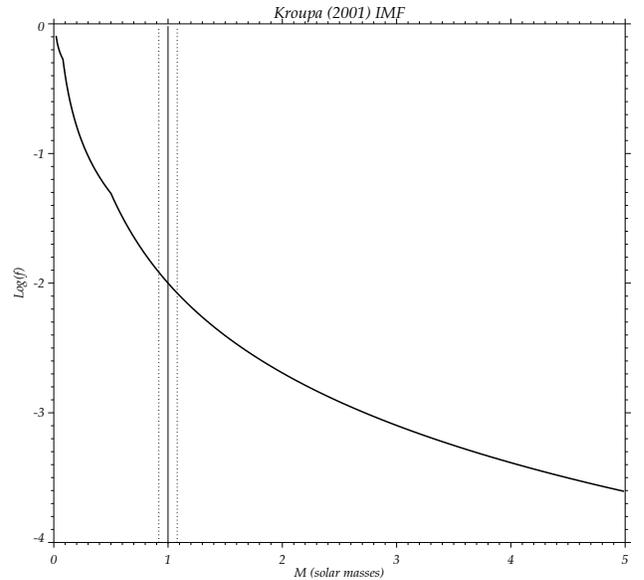}}
\caption{Kroupa (\cite{kroupa01}) galactic initial mass function with the 
low-mass cuts ($\sim 1\pm0.1\mathcal{M_{\sun}}$)  
applied for the three clusters. 
Clearly, we miss the numerous weak stars
which populate predominantly the tidal tails, observing possibly 
only 10\% of the total cluster mass.}
\label{imfkroupa}
\end{figure}

The three OCs we have chosen being nearby systems
(with a distance to the Sun in the range $\frac{1}{3}$--$\frac{2}{3}$ kpc), 
we have applied cuts in instrumental magnitudes corresponding on average to 
 about $M_V \sim 5$. 
 Using the ``universal'' mass function of Kroupa (\cite{kroupa01}), 
this implies
that we may miss about 90\% of the theoretical cluster total mass
 (see Fig.~\ref{imfkroupa}).
Note however that Scalo (\cite{scalo98}) claimed that a universal 
mass function is not justified empirically for stars with masses 
less than 1\,$\mathcal{M}_{\sun}$. Recent theoretical studies 
(Lastennet \& Valls-Gabaud~\cite{lastennet99})
also tend to show that open cluster IMFs present some variations 
(which may be due to mass segregation effects, see the discussion in Sect.~5) 
below 1.1--1.2\,$\mathcal{M}_{\sun}$. 

Therefore, estimates of the fraction of the cluster mass we observed
are accordingly subject to non-negligible variations, and 
the Kroupa (\cite{kroupa01}) IMF may rather be an 
average of possibly {\it locally} different IMFs, as already observed
even for intermediate-mass stars in OCs (e.g., Phelps \& Janes~\cite{phelps93}
or Sanner \& Geffert~\cite{sanner01}).

After this first selection from the mask in the CMD diagram 
(for example, see Fig.~\ref{n2516}) which on average keeps
$N \gtrsim$ 1500 likely members on the whole field,
we need to pay attention to a  remaining
background (Milky Way) gradient in the wide-field encompassing the OC.
 Following
Grillmair et al. (\cite{grillmair95}), as a first step we mask the cluster, 
up to 1 or 2 times its estimated tidal radius $r_{\rm t}$.
The hidden cluster is then replaced by the mean density value 
just outside it (i.e., from 1.5 to 2.5$r_{\rm t}$). 
The background $z$ is next smoothed
using a fit by a bivariate polynomial surface:
   \begin{equation}
      z(x,y) = \sum_{i}\sum_{j}{a_{ij} x^i y^j}\qquad 0 \leq i,j \leq 2.
   \end{equation}
A simple 1 $\times$ 1 bivariate polynomial is generally sufficient,
except near the Galactic Plane -- as it is often the case for OCs --
where a degree 2 is preferable in order to follow the exponential
increase of field stars towards galactic latitude zero. A
higher order polynomial is not applicable, as it may  erase
similar scale variations such as the tidal tails we are seeking. 
Then, we obtain a raw map of overdensities $Tide_{\mathrm{Raw}}$
by subtracting the Milky Way  gradient $z$ from the count of 
likely members of the cluster:

   \begin{equation}
      Tide_{\mathrm{Raw}} (x,y) = cmd(x,y) - z(x,y),
   \end{equation}
where $cmd$ is the surface density of CMD-selected stars.

At this point, we use the wavelet analysis (see Leon et al.
\cite{leon99}, \cite{leon00}). This method makes use of the so-called 
``\`a~trous'' algorithm (Bijaoui \cite{bijaoui91}). Relatively fast, it 
decomposes by a discrete wavelet transform (WT) the raw image in several 
planes $W_i (x,y)$ where $i$ is the scale of the plane, using a $B$-spline 
function. This kernel function $B_{\rm S}(x,y)$ is used for the recursive
convolutions of the initial image. If we note as $c_0$ the raw image
of overdensities $Tide_{\mathrm{Raw}}$ just obtained, we have
 successively:

   \begin{eqnarray}
      c_0 (x,y) & = & Tide_{\mathrm{Raw}} (x,y),\\
      c_i (x,y) & = & c_{i-1} \times B_{\rm S}(\frac{x}{2^i},\frac{y}{2^i});\\
      W_i (x,y) & = &  c_i (x,y) - c_{i-1}(x,y).
   \end{eqnarray}
Each of the planes $W_i (x,y)$ corresponds to a certain image
 resolution: while orders like 5 or 6 retain only ``out of
focus'' features, the lower orders give the finest details. In fact,
the last plane $W_{i{_{\rm{max}}}} (x,y)$ 
-- often known as LSP for Last Smoothed
Plane -- is not in the literal sense a true wavelet plane, but contains
the residuals of the last convolution. However, we will
consider it like the other planes, giving a total number of 7
scales. This value has proved to be a good compromise between
computing time and the maximal resolution we need.

Finally, having obtained all the  wavelet planes $W_i (x,y)$, 
we can construct the best map to recover 
overdensities around the cluster by adding several wavelet planes.
We have chosen to select only large scale planes related
to the tidal tails sizes: the best results are obtained using planes
from order 3 to the LSP (i.e., order 7 in our decomposition).
The final map of tidal tails will then be defined as in 
Leon et al. (\cite{leon00}) by the sum:
    \begin{equation}
 Tide (x,y) = \sum_{i=3}^{7} {W_i (x,y)}.  
   \end{equation}
This choice was also made in order to obtain
a good compromise between resolution and background noise.


\section{Results}
   \begin{table}
      \caption[]{Parallaxes and proper motions 
for the three selected OCs, as extracted from Baumgardt et al. 
(\cite{baumgardt00}). $N_\star$ is the number of {\sc Hipparcos}
members used to compute $\pi$ and $\mu_{\alpha^{*}\!,\delta}$. 
$\mu_{\alpha^{*}}$~includes the multiplication by the
cosine of the declination.}
         \label{ppm}
\begin{tabular}{lllll}
\hline
Cluster &  $\pi$ (mas) & $N_\star$ & $\mu_{\alpha^{*}}$ (mas/yr)& 
$\mu_{\delta}$ (mas/yr)\\ 
\hline   
NGC2287 & 1.93$\pm$0.49 &8 &  $-$4.34$\pm$0.40 & $-$0.09$\pm$0.40\\
NGC2516 & 2.77$\pm$0.25 &13& $-$4.08$\pm$0.27 & $-$10.98$\pm$0.24\\
NGC2548 & 1.63$\pm$0.79 &5&  $-$0.50$\pm$0.70 & $+$0.93$\pm$0.65\\
\hline
\end{tabular}
\end{table}

\subsection{NGC 2287 (\,$\equiv$ M41)}

This bright, intermediate-age OC was particularly
studied for  proper motions (Ianna et al. \cite{ianna87}), 
radial velocities (Amieux \cite{amieux88}) and spectrophotometric
properties (Harris et al. \cite{harris93}). 
For the present wide-field structural study
of the cluster, we found that a cut at an instrumental ($R$) mag. of 14 
offered the maximum cluster contrast with respect to
  the field. Using a  plate calibration based on the GSPC-II, it
is equivalent to $B_{\rm{cut}} \simeq 16.2$.
Given a distance to NGC 2287 of about 700 pc and the low absorption in this direction (see Table~\ref{taboc}), the cut corresponds to 
main sequence stars of about $M_V\sim 6.1$, 
hence masses of the order of 0.9\,$\mathcal{M}_{\sun}$. 
The main sequence brightest selected stars have $V\simeq $ 9,
 i.e.  mass greater than 3\,$\mathcal{M}_{\sun}$.
Taking the IMF  of Kroupa (\cite{kroupa01}) within these limits, 
our selection of members  may represent only $\approx 10\%$
of the total cluster mass.

Despite this limitation, the overall structure
observed for NGC 2287 (cf. Fig.~\ref{m41}, left) corresponds
relatively well to what is expected: a global elongated
shape nearly parallel to the Galactic Plane (i.e., from SSE to
NNW) with projected axis ratios estimated to 1.5\,:\,1.0, 
and with extensions perpendicular to it (the arrow
points perpendicularly 
towards $Z =$ 0, a clump of density is visible in this direction).
The dashed arrow indicates the proper motion ({\textsf{\textsl{PM}}},
see Tab.~\ref{ppm})
of the OC, and the dashed circle traces the cluster extension 
(radius estimated to $\sim 35'$) as given in Table~\ref{taboc}.

The ``pollution map'' (see Fig.~\ref{m41}, right) shows
no obvious clusters of galaxies and no anticorrelation between
the tidal tails and the {\sl IRAS} 100 $\mu$m map (dust absorption)
which could bias the source count. 
Otherwise, it is clearly visible in Fig.~\ref{m41} (left) 
that the star count peak is shifted by about 15$'$ 
SW from the center of the field, i.e.
the ``classical'' coordinates of NGC 2287 given in catalogues 
are incorrect.

\subsection{NGC 2516}

\begin{figure*}
\resizebox{\hsize}{!}{\includegraphics{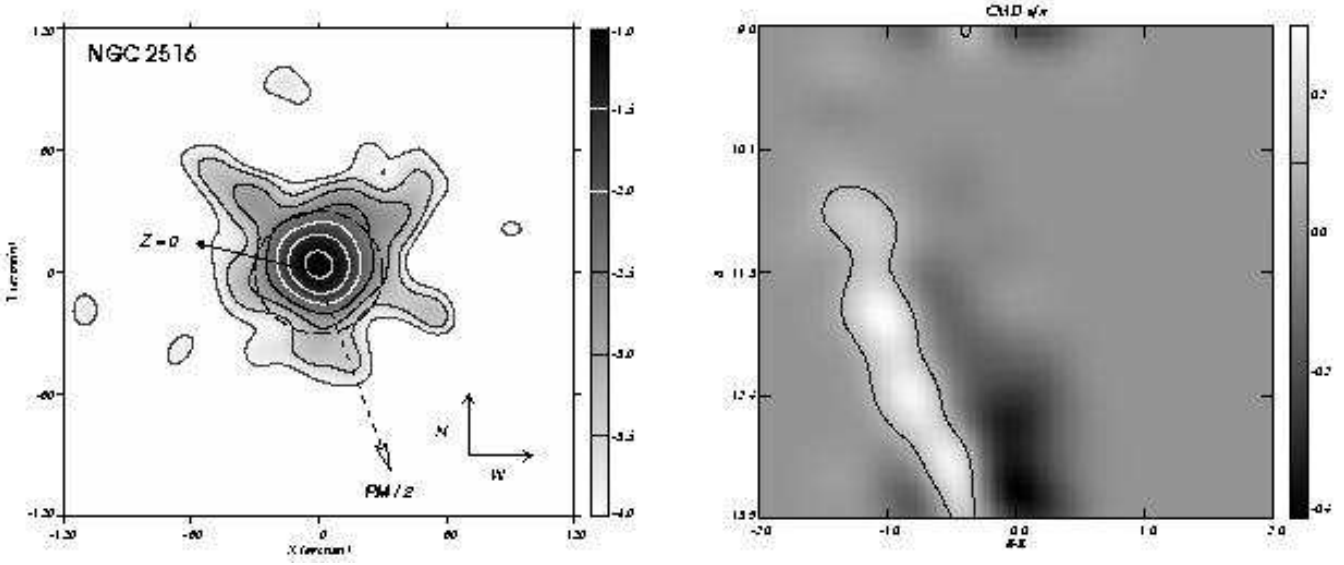}}
\caption{Left: density (Log scale) of probable NGC 2516 members, using
star-counts on a $B$--$R$ pair of Schmidt 
plates. The dashed circle represents the cluster diameter according 
to Lyng\aa\ 
(\cite{lynga87}). Dashed arrow is the proper motion ({\textsf{\textsl{PM}}}) 
of the cluster (cf. Table~\ref{ppm}) whereas the solid arrow points $\perp$ 
to the Galactic Plane towards $Z =$~0.
Right: the colour-magnitude signal to noise 
(relative to the background, see text) 
with the mask (contour) used to select
only high (bright regions) S/N areas.}
\label{n2516}
\end{figure*}

\begin{figure*}
\resizebox{\hsize}{!}{\includegraphics{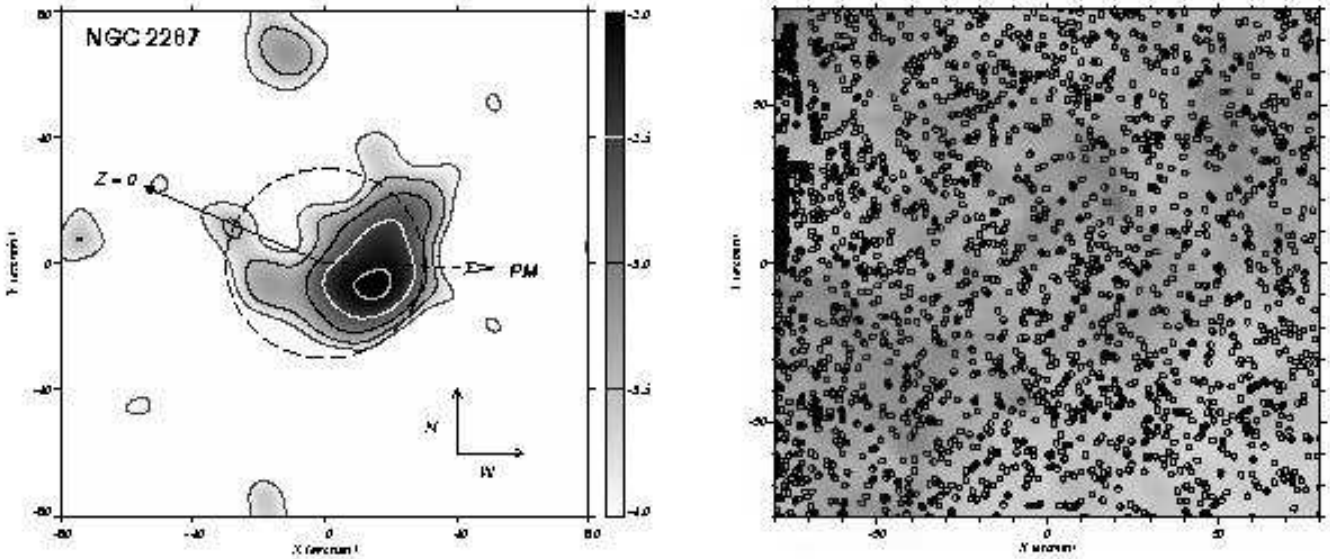}}
\caption{Density  
of NGC 2287 stars (left) and pollution map for the
same field (right). Definition of the arrows 
and dashed circle is the same as in Fig.~\ref{n2516},
 with twice the NGC 2516 
proper motion ({\textsf{\textsl{PM}}}) scale.
Right: the {\sl IRAS} 100 $\mu$m absorption is in greyscale.
Small circles represent galaxies, as extracted
from the {\textsf{SExtractor}} catalogue using a separation in a  
Log(area) {\it vs.} magnitude diagram. 
Many spurious extended source detections occur near
the E border of the plate ($X \simeq -80'$ from the cluster center).}
\label{m41}
\end{figure*}
\begin{figure*}
\resizebox{\hsize}{!}{\includegraphics{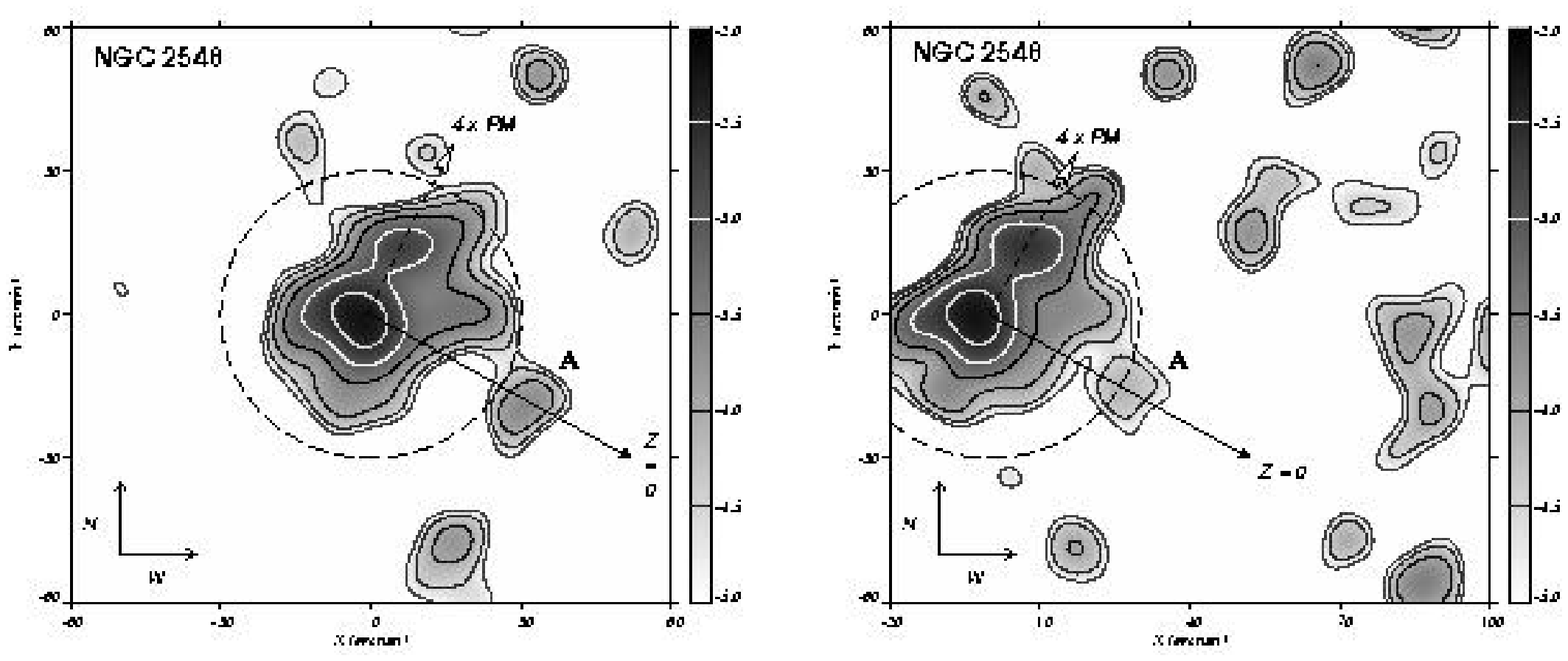}}
\caption{Field centered on NGC 2548, two pairs of $B$ \& $R$~plates.
Left: the open cluster isodensity levels on the SRC775 plates. 
Right: the center of NGC 2548 lies only 
20$'$ East of the border of the SRC776 plate. Despite this, several
features are nearly identical to the SRC775 plates (double core of the 
cluster, prolate shape and a tidal tail with a clump). 
See sect. 5 for the discussion about the clump of star {\bf A}. Proper motion 
of the OC ({\textsf{\textsl{PM}}}), 4 times enlarged, is very uncertain.} 
\label{m48}
\end{figure*}

Eggen (\cite{eggen74}) detected several members up to $\sim$5 pc 
from the cluster center and described NGC 2516 as the Southern Pleiades.
Indeed, this very bright and rich cluster shares a common age and 
$V$ absolute galactic motion with M45.  NGC 2516 contains many variable stars
and is ideal to look for low-mass and active stars.
It is also an essential cluster in the calibration of the IFMR (Initial Final
Mass Relation) as it contains several white dwarfs. 

Among the numerous studies of this spectacular OC, the one by 
Dachs \& Kabus (\cite{dachs89}) is complete up to $M_V\,=\,5\fm5$, over
a relatively wide-field. They propose a total cluster mass of about 
$\approx$\,1000\,$\mathcal{M}_{\sun}$.

An efficient  cut at an instrumental magnitude of 13.5 was done, 
as the cluster is particularly nearby ($\simeq$350~pc, see Robichon et al.
\cite{robichon99}). 
After cross-identification with the catalog 
of Dachs \& Kabus (\cite{dachs89}), this cut 
corresponds to main sequence stars brighter than $V = 12.4$ 
or $M_V \simeq 4.4$. In terms of mass,
the cut retains only stars with $\mathcal{M} 
\gtrsim 1.1\,\mathcal{M}_{\sun}$. 
The brightest CMD-selected stars have $V = 8.2$
($M_V \simeq 0.2$) or $\mathcal{M} \sim$ 2.5--3\,$\mathcal{M}_{\sun}$.
The integration of the Kroupa (\cite{kroupa01}) IMF within these limits
shows that we could miss nearly 90\% 
of the cluster potential mass, even if, as discussed in Sect. 3,
variations in the 
mass function just below our cut may significantly change
 the number of unstudied members.

Despite this low percentage in mass of selected members,
in Fig.~\ref{n2516} (left), two remarkable tails
extend up to 1.5\degr\ from the cluster center, nearly
perpendicularly to the Galactic Plane (the arrow points towards $Z=0$).
It is interesting to note the  almost perfect circular
geometry of the central parts of NGC~2516, despite the ideal location
($\ell \simeq 270\degr$) of this OC for observation of 
the predicted flattening
by the radial galactic force gradient (see the discussion in Sect. 5).

Except for a strong emission point north of the cluster (corresponding
to the ``Toby jug'' nebula IC 2220), the extinction due to dust is 
homogeneous, and no clumps in background galaxies correlate with 
the position of tails. Hence, no observational bias can
account for the presence of these huge tidal extensions.


\subsection{NGC 2548 (\,$\equiv$ M48)}

Despite being a bright cluster, NGC 2548 (later identified as the
entry \#48 in the Messier's catalogue) has not been studied recently.
Here, we have used {\it two} pairs of blue and red survey  plates, in order
to check against possible variations in the sensitivity of the
photographic emulsion which may bias the star-counts. No
such effect was found. The OC is
 situated  near the E border of the first pair (ESO/SRC 775) and
on the W border of the 2nd (ESO/SRC 776). 
Despite these locations on the plates,
the tidal tail extraction has proved consistent as shown on Fig.~\ref{m48}.

For the two pairs of plates including NGC 2548, 
we have chosen a strong cut, keeping only stars brighter than the 
instrumental magnitude 13.0 (corresponding to  about $B_{\rm{cut}} \sim$ 14.8
using a rough calibration based on GSPC-I standard stars), or $M_B\sim 5.3$.
Again, we keep only stars more massive than about 
1--1.1\,$\mathcal{M}_{\sun}$.
Interestingly, the 
inner isodensity levels show a secondary peak (double core).
The projected 
axis ratio is around 1.7\,:\,1.0. The absorption in the region of the 
OC is very low (see Table~\ref{taboc}) and uniform.
Numerous extended sources were detected but no cluster of galaxies can
account for the overall prolate shape and the SW tidal extension.

\section{Discussion}

Theoretical considerations (Wielen \cite{wielen75}) have shown that the haloes
of open clusters are flattened by the galactic gravitational field. Indeed,
the gradient of the gravitational force in the galactic radial direction
is stretching the OC: simulations (Terlevich \cite{terlevich87}; 
Portegies Zwart et al. \cite{portegies01}) confirm
that OCs take the shape of an ellipsoid with axes in
a ratio $X:Z \simeq 2.0:1.0$. The longest axis always points towards the
Galactic Center whereas the smaller one is in the $Z$ direction.
This compressed shape has been observed only in a few clusters.
In the Hyades, a flattening along the Galactic Plane was
first suspected by Oort (\cite{oort79}). Perryman et al. (\cite{perryman98})
recently confirmed, in their exhaustive study of the 3-D structure of
this fundamental ``calibrator'' cluster, the prolate shape of the
halo, which is well elongated towards the Bulge.
In the Pleiades, a similar elongated shape roughly in the $\ell$ direction
was detected by van Leeuwen (\cite{vanleeuwen83}) and definitively
established by Raboud \& Mermilliod (\cite{raboud98a}).

\begin{figure}
\resizebox{\hsize}{!}{\includegraphics{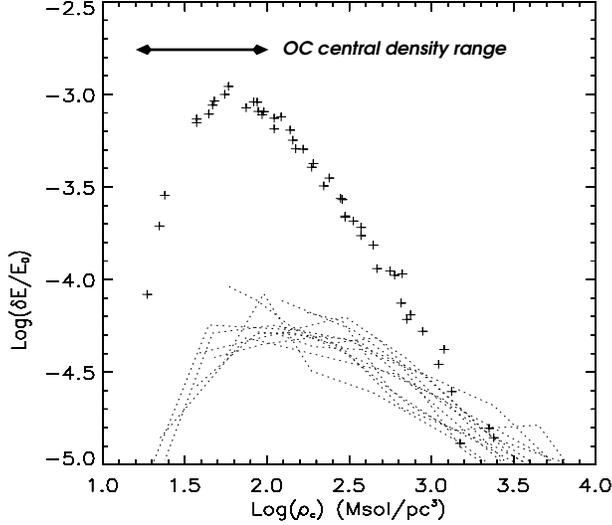}}
\caption{Contribution of the adiabatic heating during the disk crossing,
using particle perturbation simulations (Leeuwin et al. \cite{leeuwin93};
Leon \cite{leon98}). The crosses represent the gain of energy $\delta E$
 relative to the total energy
$E_0$ of the cluster. The dashed lines represent the dispersion in
$\delta E$ from different sets of simulations. We indicate the range of
rich OC central densities.}
\label{steph}
\end{figure}

It has been shown as well (Leon \cite{leon98}) 
that the disk-shocking 
compresses strongly the stellar cluster during the crossing
of the galactic Plane. In the case of OCs, the heating is
very efficient from the adiabatic component which includes the
short period stellar orbits relative to the crossing time through
the Plane. Figure~\ref{steph} shows the contribution of the
adiabatic heating which is maximum for typical open cluster
central densities $\leq100~\mathcal{M}_{\sun}$\,pc$^{-3}$ 
 (e.g., Binney \& Tremaine \cite{binney87}). 

\begin{figure*}
\resizebox{\hsize}{!}{\includegraphics{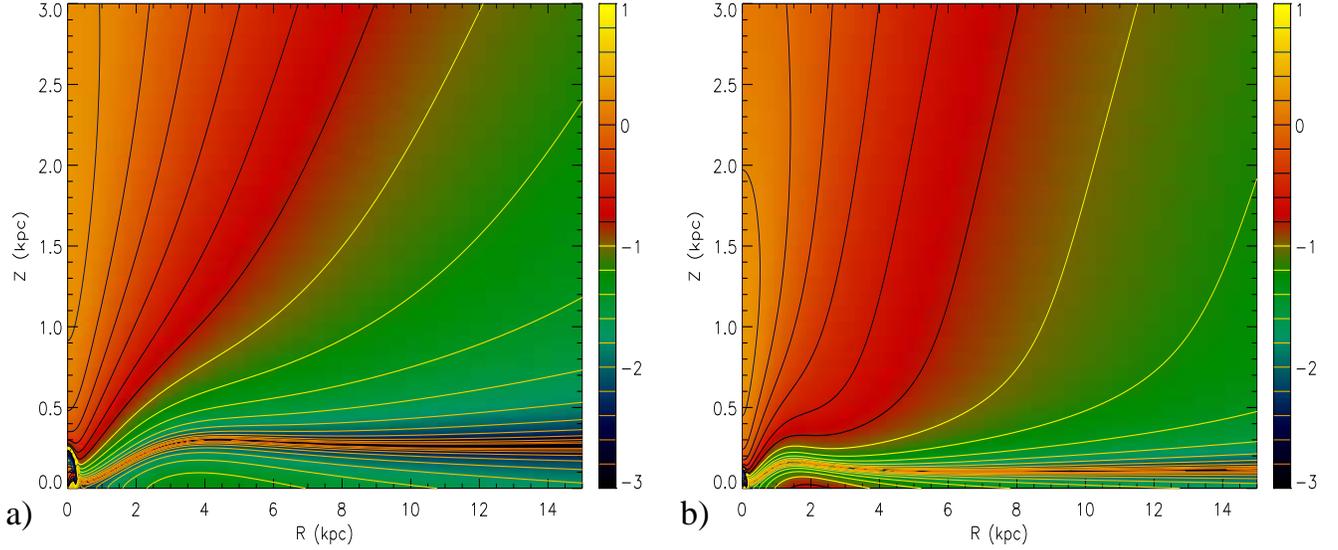}}
\caption{Ratio (Log scale) of the $Z$-tidal force to the radial tidal force, 
parallel to the Galactic 
Plane for an open cluster having a tidal radius of $r_{\rm t}= 10$ pc. 
In {\bf a)} we use the {\sf Gal--1} model 
(see Combes et al. \cite{combes99} for details)
and {\bf b)} the {\sf Gal--2} model. The main difference between 
the two models is the disk scale height, with 1 kpc for 
{\sf Gal--1} model and 400 pc for {\sf Gal--2} model. 
We did not take into account the dark matter halo.}
\label{tidal}
\end{figure*}

 To quantify the effects of
 the vertical and radial (parallel to the Disk plane)  tidal forces,
we show in Fig.~\ref{tidal} the ratio of these two tidal components 
for a rather rich OC having a $r_{\rm t}= 10$ pc tidal radius.
The galactic models are from Combes et al. (\cite{combes99}): 
the main difference is the scale height of the Disk which
is set to 1 kpc for the {\sf Gal--1} model 
and 400 pc for the {\sf Gal--2} model, 
leading to a major contribution of the 
vertical  tidal force in the latter case owing to a larger gradient due 
to the thinner disk. 

However, the vertical component of the {\it static}
tidal force is less important 
than the radial one, except on top of the Bulge where the radial 
component, in our model, is very weak. For a typical 
open cluster 
$Z$ range, the vertical/radial ratio of the tidal force components 
is between 1\% 
to 10\%. Nevertheless the shock due to the {\it time-dependent} tidal 
force will induce a major heating of the OC crossing the Plane. 
Interestingly the vertical tidal force is stronger at very low-$Z$, 
where the open cluster population is mainly 
located, to decrease to a minimum, for a constant
galactic radius, at $Z\sim 150$ pc ({\sf Gal--2} model) and 
increases towards higher $Z$ where the vertical 
component from the Bulge and the whole Disk takes over. 

Hence, in addition to the static tidal field, all OCs  
suffer repeated disk-shockings
with the galactic Plane. Indeed one typical open cluster 
at the solar radius crosses 
the Plane about 10--20 times before its dissolution in the Galaxy (see below),
only leaving an OC remnant (de la Fuente Marcos \cite{fuente98}).
For most of existing clusters, we expect to observe clumps of stars 
disrupted from the disk shocking,
in the direction perpendicular to the Galactic Plane. These clumps should have
a velocity (relative to the OC center of mass) close to the  velocity
dispersion in the cluster. In an OC of 
$\sim$500 $\mathcal{M}_{\sun}$ like the Hyades 
(evolved intermediate-age cluster, like NGC 2548),
the velocity dispersion is about 0.2--0.4 km\,s$^{-1}$ 
(de Bruijne et al. \cite{debruijne01}). 

   \begin{table}
      \caption[]{Galactic positions and heliocentric space
motions for the 3 selected
open clusters, data coming from the compilation by 
 Piatti et al. (\cite{piatti95}). $R_{\rm p}$ (resp.
$R_{\rm a}$) is the radius at the periphelion (resp.
at the aphelion) of the cluster orbit.}
         \label{uvw}
\begin{tabular}{llllllll}
 
            \hline

            Object &  $R_{\rm p}$ & $R_{\rm a}$ & $Z$ & $H^{\dag}$
& $U$ & $V$ & $W$\\
        
NGC & kpc & kpc & pc & pc & km/s & km/s & km/s\\
            \hline
          
            2287 & 7.95 & 8.97 & $-$130 & 870 & $+$25 & $+$9 & $-$80 \\
            2516 & 5.60 & 8.54 & $-$120 & 200 & $+$2 & $-$24 & $+$5 \\
            2548 & 7.43 & 8.99 & $+$160 & 420 & $+$37 & $+$5 & $-$43 \\                   
            \hline
\end{tabular}
\begin{list}{}{}
\item[$^{\dag}$] $H = |Z_{\rm{max}}|$ is the maximum height that
the open cluster is supposed to reach above (or under) the Galactic Plane. 
\end{list}
\end{table}

We can interpret the SW clump (noted {\bf A} in Fig.~\ref{m48}) around NGC~2548,
in the direction of the Plane,  as a remnant of the last disk-shocking. 
Given the distance of the clump to the center of NGC 2548 of
about $\sim$8 pc  and taking a typical velocity
dispersion of 0.3~km\,s$^{-1}$, we estimate the last shock finished $\approx$26  Myr ago. On the other hand, following Dachs \& Kabus (\cite{dachs89}), 
the period $P_Z$ of the $Z$-oscillations of an  open cluster is given by:
\begin{equation}
P_Z = 2\pi\sqrt{\frac{Z_{\rm{max}}}{K_{Z{_{\rm{max}}}}}} 
\end{equation} 
where  $K_Z$ is the galactic vertical acceleration at the maximum height 
$Z_{\rm max}$ the OC can reach. We use the acceleration values given by
Vergely et al. (\cite{vergely01}), based on recent estimates of
the disk scale height  $D = 240$ pc and 
surface mass density $\Sigma_0 = 48\,\mathcal{M}_{\sun}$\,pc$^{-2}$,
whereas the local mass density is taken as
$\rho_0 = 0.076\,\mathcal{M}_{\sun}$\,pc$^{-3}$.

The results of the compilation of the absolute 
$(U,V,W)$ motions of the three clusters by 
Piatti et al. (\cite{piatti95}) are shown in Table~\ref{uvw}. 
In the case of NCG 2548 however, different values were found recently by 
Wu  et al. (\cite{wu01}). 
These authors deduced $(U,V,W) = (1,221,3)$ km\,s$^{-1}$ from an 
exhaustive proper motion survey, with $H = |Z_{\rm{max}}| = 170$ pc
(taking the nearby distance estimate of $d_{\rm{M48}} \simeq 530$ pc, see 
Clar\'\i a \cite{claria85}).

Using the $W$ and $H$ 
values of Piatti et al. (\cite{piatti95}), we can estimate that
the last crossing of NGC~2548 through the Galactic Plane 
finished about $\sim$40 Myr ago, whereas taking the data of Wu et al.
(\cite{wu01}) the last shock occurred  $\sim$20 Myr ago. 
Both results are  consistent with our dating derived from the clump.
We note that with an age of about 360 Myr (see Table~\ref{taboc}),
NGC 2548 is a dynamically evolved OC which has already suffered 
about 7 or 8 disk-shockings. 

It has also been shown (Leon \cite{leon98}; Combes et al. \cite{combes99})
that during the disk-shocking the shape of the
cluster will change rapidly before the relaxation, which is never
reached for an OC constantly oscillating in the Plane. The difference in
shape between NGC~2516 (round core) and both NGC~2287 \& NGC~2548 (pronounced
prolate shape) can be accounted for a transient phase during the shock.
 It could also be due to effects of encounters 
with passing molecular clouds which can reduce, 
or randomize the orientation of the flattening (see Theuns \cite{theuns92a},
\cite{theuns92b}). 

In order to estimate the importance of the disk-shocking in the 
evolution of the OCs, we have performed $N$-body simulations. 
We used the so-called perturbation particles  method 
(Leeuwin et al. \cite{leeuwin93}).
To solve the collisionless Boltzmann equation, this technique combines an 
analytical description of
the equilibrium state and a numerical evaluation of the 
local density perturbation. Indeed perturbation particles are well suited 
for a collisionless system, and can be used for simulating a disk-shocking
because
the relaxation time of the cluster is still larger than the crossing time. 
The simulated cluster is modeled by a Plummer's sphere for which the 
gravitational potential and the distribution function are 
known analytically (e.g., Leon \cite{leon98}).

We stress that our simulations are only aimed at giving an estimation 
of the mass loss. In particular, we have not included the
 effect of binarity and mass segregation. Binary stars, primordial or 
exchanged, seem to be less important for the evolution of rich clusters 
like those 
 considered in our models (see de la Fuente Marcos~\cite{fuente96}) 
and may be regarded as more massive stars, hence defering the problem 
to that of mass segregation. 

Furthermore, the
perturbation particles method does not take into account the mass 
segregation process, but during one crossing this latter will be slow
(see Combes et al.~\cite{combes99}). The spatial mass function is
certainly  different between the core and the outskirts of the cluster 
(e.g., Raboud \& Mermilliod \cite{raboud98b}). However, 
during a crossing the tidal effects are independent of the stellar masses.
As a result, ignoring the mass segregation is not expected to strongly
bias the {\it total} mass of the tidal tail
(only the mass function will
be different, with an excess of low mass stars).

\begin{figure}
\resizebox{\hsize}{!}{\includegraphics{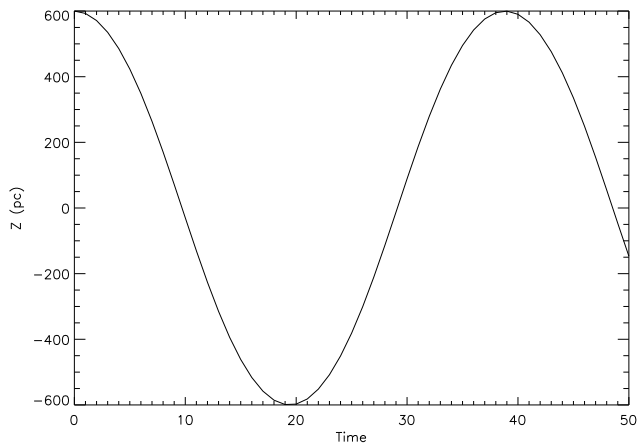}}
\caption{$Z$ position of the open cluster {\it vs.} the time 
(expressed in $N$-body time units) in the perturbation particle simulation.}
\label{pos_simul}
\end{figure}

\begin{figure}
\resizebox{\hsize}{!}{\includegraphics{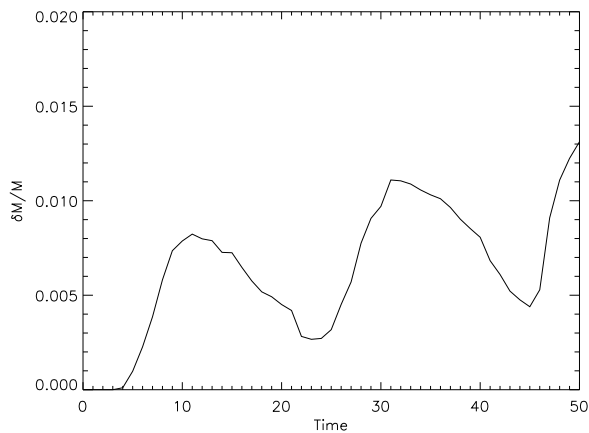}}
\caption{
Mass loss relative to the total mass of the cluster {\it vs.}
 time ($N$-body units) in the perturbation particle simulation.}
\label{massloss}
\end{figure}

We have followed an OC starting at 600 pc (see Fig.~\ref{pos_simul}) 
above the Plane at the solar galactic radius
with a total mass of 1000\,$\mathcal{M}_{\sun}$ and
a core radius  $r_{\rm c}$ of  1 pc. We have chosen such a high-$Z$ cluster 
to avoid strong gradient at the beginning of the simulation.
Old ``survivors'' like M67 (Fan et al. \cite{fan96}) 
or NGC 188 (Sarajedini et al.
\cite{sarajedini99}) have total masses of this order, and follow such 
high-$Z$ orbits (see Carraro \& Chiosi \cite{carraro94}).  
Figure~\ref{massloss} shows that the OC mass loss by the disk-shocking 
is about 1.5\% of its total mass in one $Z$-oscillation. 
It can be noticed that some stars are re-bound to the cluster
after the crossing. 

Despite the lower intensity of the 
static $Z$-tidal force relative to the radial one, the {\it time-dependent} 
$Z$-tidal force during the crossing strongly speeds up the destruction 
of the OC. The global contribution of the disk-shocking can be estimated
to be about 10 to 20\% for such a high-$Z$ cluster, 
and likely more for a younger open cluster which experiences 
stronger $Z$-force gradients in the vicinity of the Plane (see above).
We can try to relate this mass-loss to the mean life duration of 
open clusters in the Galaxy. 

\begin{figure}
\resizebox{\hsize}{!}{\includegraphics{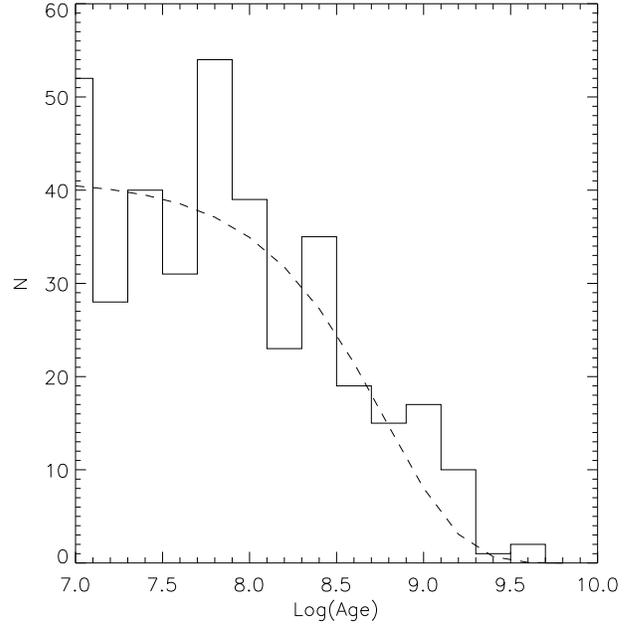}}
\caption{Distribution of the number of OCs {\it vs.} the age (Log) 
from the catalogue of Loktin et al. (\cite{loktin01}).
 The exponential fit, excluding the youngest and the
oldest Galactic open clusters, with a time-scale of 612 Myr is
overplotted as a dashed line.} 
\label{vieoc}
\end{figure}

Using a pre-release of the catalogue of Loktin et al. (\cite{loktin01}) 
available on the {\it Base des Amas} web database, we show in Fig.~\ref{vieoc} the
age distribution of the  423 open clusters in their sample.
As mentioned above, OCs suffer different destructive 
processes (disk-shockings,
encounters with GMCs, tidal field harassment) 
which lead to the rapid
depletion of the whole population. It appears that the population age, 
excluding both very young  ($\leq 10^7$\,yr) and very old ($\geq 1$ Gyr) OCs,  
can be modeled by an exponential law of the following form 
(see the fit in Fig.~\ref{vieoc}):
\begin{equation}
N=N_\star \ \rm{e}^{-\frac{\tau}{\tau_\star}},
\end{equation}
where $\tau_\star$ is a time-scale for OCs (primarily in the solar
neighbourhood, as catalogues become incomplete for distant clusters)
which corresponds to their destruction time-scale.
Assuming a constant star-formation rate over the last 10$^9$ years
in the solar neighbourhood (e.g., Rocha-Pinto et al. \cite{rocha00}), 
the OC destruction time-scale 
is of the order of $6\times10^8$ yr, 
somewhat superior to previous observational 
estimates (Lyng\aa\ \cite{lynga82}; Janes \& Phelps \cite{janes94}).
Selection effects in different Galactic OCs catalogues may explain
this small discrepancy. Interestingly, the value of 600 Myr is
in good agreement with the OC disintegration time in the field
$\tau{_{_{0.1}}}$
defined by Kroupa (\cite{kroupa95}) in its realistic simulations.

As stated before, in the
solar neighbourhood, disk shocking seems to be very important in the
destruction of OCs. An OC at the solar radius suffers about ten to twenty
crossings during this time-scale, which leads to an upper-limit for the
mass loss efficiency during one disk-crossing  between 5 to 10\%. This value
is clearly more important than for globular clusters (Leon \cite{leon98})
 which are much more massive and concentrated stellar aggregates.

\section{Conclusions}
We have detected important tidal extensions around 3 bright open clusters
using photometric selection on wide-field photographic plates. 
Density isophotes of cluster probable members go further away from the
ordinary limits given for these objects. In two cases (NGC~2287 \& NGC~2548) 
we have observed strong flattening of the cluster
that we interpret  as a consequence of the disk-shocking. 
We stress that the adiabatic heating is important for the 
open cluster evolution, as well as the time varying galactic field.
In spite of the weakness of the static $Z$-tidal force field, the time varying
tidal field during the disk-shocking will indeed affect deeply the dynamical
evolution of OCs. Its effects, computed with the perturbation
particles method, account for about 20\% of the mass loss.
We also estimate the destruction 
time-scale of open clusters in the solar neighbourhood to 600 Myr.
In the case of NGC 2548, it has been
possible to estimate that the last shocking with the Galactic 
Plane occurred about 30 Myr ago.   
This value should be confirmed by a complete proper motion 
and radial velocities study.  
Advent of super-MOS (e.g. {\sl Flames}) and, at medium-term, {\sc Gaia} will 
provide hundreds of $V_{\rm r}$ and very accurate proper 
motions even for low-mass members of known OCs.
Moreover, accurate multi-band, deep wide-field CCD photometry 
of a sample of open clusters will provide much better selection of members,
including sub-solar mass stars.  
The comparison of these new observations with more realistic simulations
 will be the subject of forthcoming papers.

\begin{acknowledgements}
Special
 thanks to R. Chesnel and P.~Toupet for plate scanning and pre-reduction.
S. L. was supported in part by the Deutsche Forschungsgemeinschaft (DFG)
via grant SFB 494, by special funding from the Science Ministry of the Land
Nordrhein-Westfalen. We gratefully acknowledge F. Combes and F. Leeuwin for
invaluable discussions. Many thanks to O. Bienaym\'e and Wu et al. 
for having kindly provided their results before publication.
Finally, we are indebted to an anonymous referee for his/her suggestions 
and  comments which helped to improve this paper.
\end{acknowledgements}



\begin{thebibliography}{}

\bibitem[1988]{amieux88}
Amieux, G. 1998, A\&AS, 76, 305

\bibitem[1998]{baumgardt98}
Baumgardt, H.    1998, A\&A, 340, 402

\bibitem[2000]{baumgardt00}
Baumgardt, H., Dettbarn, C., \& Wielen, R. 2000, A\&AS, 146, 251

\bibitem[1991]{berger91}
Berger, J., Cordoni, J.-P., Fraingant, A.-M., Guibert, J., Moreau, O.,
Reboul, H., \& Vanderriest, C. 1991, A\&AS, 87, 389

\bibitem[2001]{bergond01}
Bergond, G.,  Guibert, J., Leon, S., \& Vallenari, A. 2001, in ASP~Conf.
Ser. 228, Dynamics of Star clusters and the Milky Way, ed. 
S. Deiters, B. Fuchs, A. Just, R.  Spurzem, \& R. Wielen, 386, in press 

\bibitem[1996]{bertin96}
Bertin, E., \& Arnouts, S. 1996, A\&AS, 117, 393

\bibitem[2001]{bica01}
Bica, E., Santiago, B. X., Dutra, C. M.,
 Dottori, H., de Oliveira, M. R., \& Pavani, D. 2001, A\&A, 366, 827

\bibitem[1991]{bijaoui91}
Bijaoui, A. 1991, in  3rd ESO/ST-ECF Data Analysis Workshop, Garching, 
The Wavelet Transform, 3

\bibitem[1987]{binney87}
Binney, J., \& Tremaine, S. 1987, Galactic Dynamics (Princeton University Press)

\bibitem[1983]{bruch83}
Bruch, A., \& Sanders, W. L. 1983, A\&A, 121, 237

\bibitem[2001]{debruijne01}
de Bruijne, J. H. J.,
 Hoogerwerf, R., \& de Zeeuw, P. T.     2001, A\&A, 367, 111

\bibitem[1994]{carraro94}
 Carraro, G., \& Chiosi, C. 1994, A\&A, 288, 751

\bibitem[1998]{chen98}
 Chen, B., Vergely, J.-L., Valette, B., \& Carraro, G. 1998, A\&A, 336, 137 

\bibitem[1985]{claria85}
Clar\'\i a, J. J. 1985, A\&AS, 59, 195

\bibitem[1999]{combes99}
Combes, F., Leon, S., \& Meylan, G. 1999, A\&A, 352, 149

\bibitem[1989]{dachs89}
Dachs, J., \& Kabus, H. 1989, A\&AS, 78, 25 

\bibitem[1974]{eggen74}
Eggen, O. J. 1974, ApJ, 188, 59

\bibitem[1996]{fan96}
Fan, X., Burstein, D., Chen, J.-S.,  et al. 
1996, AJ 112, 628

\bibitem[1996]{fuente96}
de la Fuente Marcos, R. 1996, A\&A, 314, 453

\bibitem[1997]{fuente97}
de la Fuente Marcos, R. 1997, A\&A, 322, 764

\bibitem[1998]{fuente98}
de la Fuente Marcos, R. 1998, A\&A, 333, L27

\bibitem[1995]{grillmair95}
Grillmair, C. J., Freeman, K. C., Irwin, M., \& Quinn, P. J. 1995, AJ, 109, 2553

\bibitem[1993]{harris93}
Harris, G., Fitzgerald, M., Mehta, S., \& Reed, B. 1987, AJ, 106, 1533

\bibitem[1987]{ianna87}
Ianna, P. A., Adler, D. S., \& Faudree, E. F. 1987, AJ, 93, 347

\bibitem[1982]{janes82}
Janes, K. A., \& Adler, D. S.   1982, ApJS, 49, 425

\bibitem[1994]{janes94}
Janes, K. A., \& Phelps, R. L. 1994, AJ, 108, 1773

\bibitem[1997]{jeffries97}
Jeffries, R. D., Thurston, M. R., \& Pye, J. P. 1997, MNRAS, 287, 350

\bibitem[1995]{kroupa95}
Kroupa, P. 1995, MNRAS, 277, 1522

\bibitem[2001]{kroupa01}
Kroupa, P. 2001, MNRAS, 322, 231

\bibitem[1983]{vanleeuwen83}
van Leeuwen, F. 1983, PhD dissertation (Sterrenwacht Leiden)

\bibitem[1999]{lastennet99}
Lastennet, E., \& Valls-Gabaud, D. 1999, Rev. Mex. Astron. Astrofis., 8, 115

\bibitem[1993]{leeuwin93}
Leeuwin, F., Combes, F., \& Binney, J. 1993, MNRAS,  262, 1013

\bibitem[1998]{leon98}
Leon, S., 1998, PhD dissertation (Observatoire de Paris)

\bibitem[1999]{leon99}
Leon, S., Bergond, G., \& Vallenari, A. 1999, A\&A, 344, 450

\bibitem[2000]{leon00}
Leon, S., Meylan, G., \& Combes, F. 2000, A\&A,  359, 907

\bibitem[2001]{loktin01}
Loktin, A. V., Gerasimenko, T. P., \& Malisheva, L. K. 2001, 
A\&AT, in press

\bibitem[1982]{lynga82} 
Lyng\aa, G. 1982, A\&A, 109, 213

\bibitem[1987]{lynga87} 
Lyng\aa, G. 1987, Lund Catalogue of Open Clusters 
(5th edition, CDS Strasbourg) 

\bibitem[1995]{mermilliod95}
Mermilliod, J.-C. 1995, in Information \& On-Line Data in Astronomy,
Eds. D. Egret and M. A. Albrecht (Kluwer Academic Press),  127 

\bibitem[1993]{meynet93}
Meynet, G., Mermilliod, J.-C.,  \& Maeder, A. 1993, A\&AS, 98, 477

\bibitem[1998]{odenkirchen98}
Odenkirchen, M., Soubiran, C., \& Colin, J.    1998, NewA, 3, 583

\bibitem[1979]{oort79}
Oort, J. H. 1979, A\&A, 78, 312

\bibitem[1987]{pandey87}
Pandey, A. K., Bhatt, B. C., \& Mahra, H. S. 1987, Ap\&SS, 129, 293

\bibitem[1987]{pandeymahra87}
Pandey, A. K.,  \& Mahra, H. S. 1987, MNRAS, 226, 635

\bibitem[1998]{perryman98}
Perryman, M. A. C., Brown, A. G. A., Lebreton, Y., G\'omez, A., Turon, C.,
Cayrel de Strobel, G., Mermilliod, J.-C., Robichon, N., Kovalevsky, J.,
\& Crifo, F. 1998, A\&A, 331, 81
	
\bibitem[1993]{phelps93}
Phelps, R. L., \& Janes, K. A. 1993, AJ, 106, 1870

\bibitem[1995]{piatti95}
Piatti, A. E., Clar\'\i a, J. J., \& Abadi, M. G. 1995, AJ, 110, 2813

\bibitem[2001]{portegies01}
Portegies Zwart, S., McMillan, S. L. W., Hut, P., \& Makino, J. 2001,
MNRAS, 321, 199

\bibitem[1998a]{raboud98a}
Raboud, D., \& Mermilliod, J.-C. 1998, A\&A, 329, 101

\bibitem[1998b]{raboud98b}
Raboud, D., \& Mermilliod, J.-C. 1998, A\&A, 333, 897


\bibitem[1999]{robichon99}
Robichon, N., Arenou, F., Mermilliod, J.-C., \& Turon, C. 1999, A\&A,  345, 471

\bibitem[2000]{rocha00}
Rocha-Pinto, H. J., Scalo, J., Maciel, W. J., \& Flynn, C. 2000, A\&A, 358, 869

\bibitem[2001]{sanner01}
Sanner, J., \& Geffert, M. 2001, A\&A, 370, 87

\bibitem[1999]{sarajedini99}
Sarajedini, A., von Hippel, T., Kozhurina-Platais, V., \& Demarque, P.  1999,
 AJ, 118, 2894

\bibitem[1998]{scalo98}
Scalo., J. M. 1998, in ASP~Conf. Ser. 142, The IMF revisited -- 
A case for variations, ed. G. Gilmore, \& D. Howell, 201

\bibitem[1987]{terlevich87}
Terlevich, E. 1987, MNRAS, 224, 193

\bibitem[1992a]{theuns92a}
Theuns, T. 1992a, A\&A, 259, 493

\bibitem[1992b]{theuns92b}
Theuns, T. 1992b, A\&A, 259, 503

\bibitem[2001]{vergely01}
Vergely, J.-L, Freire, R., Egret, D., \& Bienaym\'e, O. 2000, A\&A submitted
 
\bibitem[1975]{wielen75}
Wielen, R. 1975, in Dynamics of stellar systems, ed. A. Hayli 
(Dordrecht),  119

\bibitem[1991]{wielen91}
Wielen, R. 1991, in ASP Conf. Ser. 13, The Formation and Evolution of Star Clusters, ed.  K. Janes, 343

\bibitem[2001]{wu01}
Wu, Z. Y., Tian, K. P., Balaguer-N\'u\~nez, L., Jordi, C., Zhao, J. L., \&
Guibert, J. 2001,  A\&A, submitted

\end{thebibliography}
\end{document}